# A Mechanical Study of a Glass Fabric-Thermoplastic Resin Composite: 3D-DIC and X-ray tomographic observations explained by numerical simulations based on a spectral solver.

**Authors**: Zakariya Boufaida, Julien Boisse, Stéphane André, Laurent Farge

Laboratoire d'Energétique et de Mécanique Théorique et Appliquée
CNRS UMR 7563
2, avenue de la Forêt de Haye
TSA 60604 - 54518 Vandoeuvre-lès-Nancy Cedex France

Corresponding authors: julien.boisse@univ-lorraine.fr, stephane.andre@univ-lorraine.fr

**ABSTRACT:**

In the study presented in this paper, we analyzed the mechanical response of a glass fiber plain weave/polymer composite at the fabric millimetric mesoscale. The detail of the stress and strain fields in a fabric repeating unit cell was numerically calculated using CraFT (**C**omposite **r**esponse **a**nd **F**ourier **T**ransforms), a code specifically conceived for simulating the mechanical behaviour of materials with complex microstructure. The local strain fields obtained by simulation were found to be in very good agreement with measurements carried out using 3D Digital Image Correlation (3D DIC). From numerical stress fields calculated with the CraFT solver, we also highlighted the subregions inside the periodic mesostructure where there is maximum stress. Furthermore, with X-ray tomography post mortem measurements, we were able to confirm that certain damage modes were well initiated in these microstructure subregions of stress concentration.

## 1. Introduction

Industry's widespread use of composite materials by the industry is still of paramount importance especially in the field of transport for evident reasons linked to structure alleviation, fuel economy and environmental concerns. Nowadays composite materials are not just used for the fabrication of accessories (appearance parts) for vehicles and enter into the design and production of structural parts (car industry) and high performance parts (aeronautics). This shift brings additional constraints and thus more research is continually required into how to produce better predictive calculations. One major research theme driven by a permanent search for innovation derives from the multiple possibilities of selecting the basic materials for the building blocks of a composite material:
- Different reinforcement materials - either technical or bio-sourced,
- Different armor architectures (e.g. types of fabrics, laminated or 3D textile structural composites...),
- Wide range of possible materials for the matrix, especially in the present context of replacing thermoset materials by thermoplastics for recycling purposes,
- Different techniques to modify the reinforcement material-matrix adhesion which is a key point for the global mechanical performance.

To analyze the influence of the internal structure of the material on the overall mechanical properties, it is essential to take into consideration both averaged fields and full local fields. For instance, damage and fracturing are heavily dependent on the local details of stress or strain fields. It is thus essential to possess reliable and efficient tools for this which are also able to produce calculations systematically, moving easily from one type of architecture to another and correctly taking the structure at different scales into account. All of this means that modelling issues in this field are increasingly important.

Initially, analytical homogenization approaches were favoured to predict macroscopic equivalent material properties. Then, the joint development of powerful computers and open-source or commercial software for simulation engineering has led to a majority of studies on composites taking up the challenge through discretization methods of the Finite Elements type [Sierles et al., 2001, Yu et al., 2002, Barbero et al., 2006, Komeili and Milani, 2012, Piezel et al., 2012, Liu et al., 2013, Jia et al., 2013]. These tools have one important limitation - obtaining highly resolved calculations requires very thin meshing in multiple areas which results in very high computational costs. An alternative has been developed over the past two decades. The CraFT code (Composite response and Fourier Transforms [http://craft.lma.cnrs-mrs.fr/, Moulinec and Suquet, 1994, 1998]) and the DAMASK code (Düsseldorf Advanced MAterial Simulation Kit [http://damask.mpie.de/, Roters et al, 2012, Eisenlohr et al., 2013]) were developed to avoid this difficulty by using a spectral approach. These two codes rely on the use of 3D Fourier Transforms and are very well-suited to modelling the behavior of complex materials at the bulk scale where a periodic pattern (Representative Elementary Volume - REV) can be found. These types of spectral methods have also proved very efficient in other fields of engineering sciences like heat transfer [Maillet et al., 2001]. Regarding the study presented here of plain weave reinforced composites, these spectral solvers give very precise and fast simulations of the 3D behavior of a clearly defined macroscopic REV. Strictly speaking, they do not require any meshing and produce highly resolved simulations of materials presenting very strong heterogeneities of properties and scales. Our research led us to prefer the CraFT code developed by Moulinec and Suquet (1994,1998). The authors initial goal was to develop an efficient method for numerically solving the local equilibrium equations in the case of complex heterogeneous materials whose microstructure could be obtained directly from experimental imaging (microscopy, tomography…). The method was first applied in the framework of elasticity and extended to the case of materials with arbitrary phase contrast and elasto-plastic and linear hardening behavior [Michel et al., 2011]. It remains nevertheless restricted to small transformations (kinematic framework of infinitesimal strains).

The aim of this work is to present a specific study of a glass fiber plain weave fabric / thermoplastic resin composite which provides mechanical modelling of the whole material in a quasi-static test and compares the results in terms of strain field measurements obtained with 3D-Digital Image Correlation (DIC). Most of the experimental features and results on this material (including fatigue analysis and mechanical spectroscopy) have been explained and reported elsewhere [Boufaida, 2015] and are only used here to validate computational results. As the results are particularly convincing, simulations of the associated stress fields will also be shown to enhance knowledge and analysis of damage phenomena which is proved consistent with post-mortem observations yielded by X-ray tomography. One important point to stress regarding the study is that it concerns in-plane shear loading as this allows the intensification of solicitations at the fiber/matrix interfaces.

The paper is structured as follows. In section 2, the modelling tool will be presented. Results of the simulation in terms of strain fields will be given in Section 3 and discussed in relation with their experimental counterparts. Section 4 will focus on predicted stress fields to investigate the damaging aspect of this material, especially regarding the role of the matrix-strand debonding. We summarize our findings and give ideas regarding future prospects in section 5.

## 2. Spectral solver for composite simulations

For the well-information of the reader about the spectral approach and its conceptual simplicity, we shall first recall the basic mathematical principles applied to the local equilibrium problem. More details can be found in Moulinec and Suquet (1998). Numerical simulations provided by CraFT and DAMASK for the same test-case will be given to show how well the algorithm works. Lastly, the model microstructure based on the reproduction of real specimen will be detailed along with explanations about how to reproduce it numerically and the associated material properties.

*2.1 Numerical method*

On a REV $V$, the local problem consists of equilibrium equations (1-a), behavior laws for the different phases (1-b) and boundary and interface conditions (1-c). For the whole study, perfectly bonded interfaces between all phases are considered (continuous evolution of the displacement and stress variables). The boundary conditions were assumed to be periodic which is in phase with the modelling of a REV as a repeated pattern of a given microstructure. All of this gives

$$div(\boldsymbol{\sigma}) = 0 \text{ on } V \quad (1\text{-a})$$

$$\boldsymbol{\sigma}(x) = c(x) : \varepsilon(u(x)) = c(x) : \varepsilon(u^*(x) + Ex) \quad (1\text{-b})$$

$$u^* \#, \boldsymbol{\sigma} \cdot n - \# \quad (1\text{-c})$$

where $c(x)$ is the rigidity tensor. As a function of space, it was indeed specified through the values of the mechanical properties associated to each "pixel" of the numerical image of the microstructure (having multiple images in the third direction to create the 3D structure). Additionally, eq.(1-b) shows that the strain field is decomposed into its average $\langle \varepsilon(u(x)) \rangle_x = E$ corresponding to the prescribed strain onto $V$ and a fluctuation term $\varepsilon(u^*(x))$. Eq.(1-c) points out that the fluctuating term $u^*$ is assumed to be periodic (symbol $\#$) while the local stress at the boundaries is anti-periodic (symbol $-\#$).

To solve this problem, an auxiliary problem is defined, based on the assumed knowledge of a homogeneous, isotropic and linear elastic reference material whose rigidity tensor is known as $c^0$. The equation below can then be substituted for eq. (1-b)

$$\boldsymbol{\sigma}(x) = c^0 : \left(\varepsilon(u^*(x)) + E\right) + \tau(x) \quad (1\text{-b})$$

with $\tau(x) = \delta c(x) : (\varepsilon(u^*) + E)$ a priori unknown, and $\delta c(x) = c(x) - c^0$.

In practical terms, the reference tensor $c^0$ is determined by the user and affects the convergence rate of the algorithm. For a two-phase composite material for example, it can be initialized with the average of the Lamé coefficients on both phases.

In a second step, the solution of (1) is expressed in real and Fourier spaces. In the real space, considering the Green operator $\Gamma^0(x)$ verifying the null divergence of the stress tensor along with periodic boundary conditions and homogeneous elastic tensor $c^0$, the complete strain solution of eq.(1) can be obtained by linearity property, as

$$\varepsilon(u(x)) = -\Gamma^0(x) * \tau(x) + E \qquad (2)$$

Its mirror counterpart in Fourier space (notation $\hat{\ }$ and transformed variable $\xi$) is

$$\hat{\varepsilon}(\xi) = -\hat{\Gamma}^0(\xi) : \hat{\tau}(\xi) \quad \forall \xi \neq 0, \quad \hat{\varepsilon}(0) = E \qquad (3)$$

The way the Green operator is computed can be found in many textbooks [T.Mura, 1987]. This is done directly in the Fourier domain (see Appendix A of Moulinec and Suquet, 1998 for further details). The solution to problem (1) then needs to be recast into the solution of equation (4) known as the *Lippman-Schwinger periodic* equation,

$$\begin{cases} \varepsilon(\mathbf{u}) = \mathbf{E} - \Gamma^0 * (\delta \mathbf{c} : \varepsilon(\mathbf{u})) \\ \hat{\varepsilon}(\xi) = -\hat{\Gamma}^0(\xi) : (\delta \mathbf{c} : \varepsilon)(\xi) \quad \forall \xi \neq 0, \quad \hat{\varepsilon}(0) = \mathbf{E} \end{cases} \qquad (4)$$

which can be solved numerically within a fixed-point algorithm and by performing the jumps between real and Fourier spaces with FFT algorithms available in mathematical libraries. The algorithm follows the path presented below (with $\mathcal{F}^{-1}$ and $\mathcal{F}$ denoting inverse and direct Fourier Transform).

The convergence test is based on the decrease and transition below a prescribed value of a relative error defined as to ensure that equilibrium is reached. In the computing code, this algorithm applies to the discretized coordinates of pixels in real space and to the corresponding discretized frequencies in Fourier space.

$$\begin{cases}
\underline{\textit{Initialization}:} \quad \varepsilon^0(\mathbf{x}) = \mathbf{E}, \quad \forall \mathbf{x} \in V \\
\underline{\textit{Iterate} \quad i+1:} \quad \varepsilon^i \quad \text{known} \\
a) \quad \sigma^i = \mathbf{c}(\mathbf{x}):\varepsilon^i \ ; \quad \hat{\sigma}^i = \mathcal{F}(\sigma^i) \quad \textit{Convergence test} \mapsto \textit{exit}? \\
b) \quad \tau^i = \sigma^i - \mathbf{c}^0:\varepsilon^i \\
c) \quad \hat{\tau}^i = \mathcal{F}(\tau^i) \\
d) \quad \hat{\varepsilon}^{i+1}(\xi) = -\hat{\Gamma}^0(\xi):\hat{\tau}^i \quad \forall \xi \neq \mathbf{0}, \quad \hat{\varepsilon}^{i+1}(\mathbf{0}) = \mathbf{E} \\
e) \quad \varepsilon^{i+1} = \mathcal{F}^{-1}(\hat{\varepsilon}^{i+1}) \\
f) \quad i \leftarrow i+1
\end{cases}$$

The CraFT code has been improved over the past ten years and validated for many different test cases. This study provides new validation of the code by comparing CraFT results with the results obtained with the DAMASK code. The test case consists of the simulation of a periodic unit cell containing 64 randomly-placed circular fibers (Figure below), subjected to a constant (average) deformation rate corresponding to the simulation of the woven tow behavior of the fabrics used for composites. Calculations were performed in the pure elastic case (Figs 1-a, 1-b) and in the case of a matrix assumed to be plastic with linear hardening (Figs 1-c, 1-d). Exactly the same results were obtained. Very small differences in the equivalent strain distribution can be seen by a close look at Figs. 1a-1b and 1c-1d. However at the meso-scale, in terms of the resulting stress-strain curves, the agreement is perfect as is shown by Figure 2 which corresponds to the case of Figs. 1-cd, where the matrix is considered to have elastic-plastic behavior with isotropic linear hardening. In terms of computational efficiency, the above simulations required a CPU time of 2000s to 4700 s (depending respectively on whether the pure elastic or elasto-plastic case was processed) on 2 over 8 cores of a standard computer (Intel® Xeon(R) CPU W3520 @ 2.67GHz $\times$ 8 ).

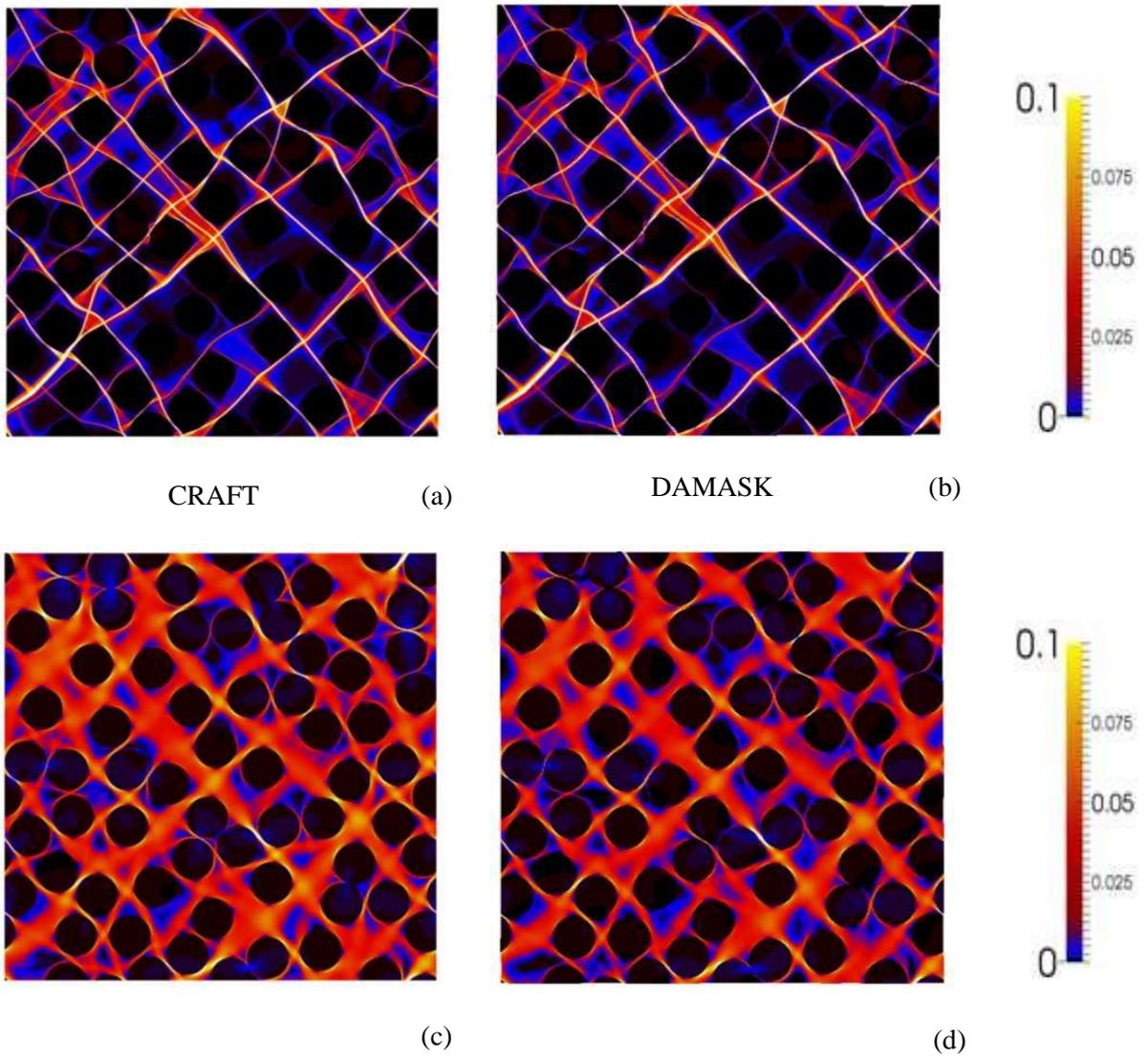

Figure 1 : Simulations of the equivalent deformation field of a woven tow with randomly distributed circular fibers using CRAFT code (a - c) and DAMASK code (b - d).

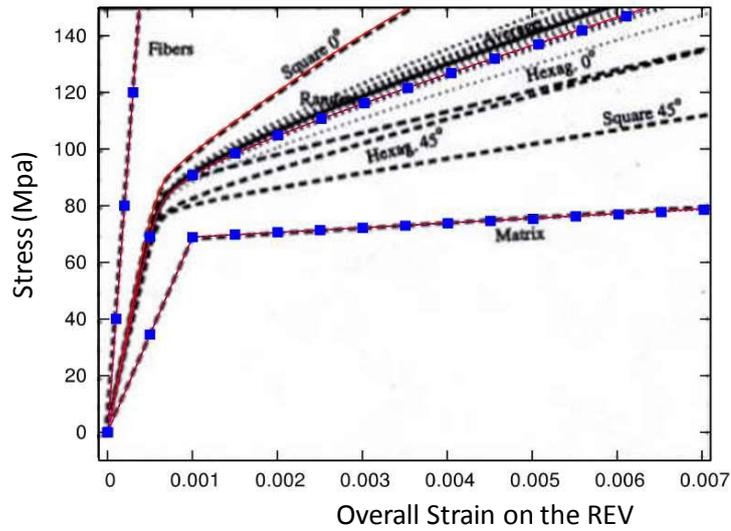

Figure 2: Comparison of the results (i) published in Moulinec and Suquet (1998, Fig.6), (ii) obtained by CRAFT for the random arrangement shown in Fig.1 (solid line), (iii) obtained by DAMASK for the same arrangement (square dots).

*2.2 Input microstructure model*

The composite material under study is modelled at the mesoscale, i.e. the scale corresponding to the internal structure of the fabric which structures the mechanical stress/strain fields. In all following figures, the (XY) or (12) planes correspond to the plane of the plain weave lamina. Thus the through-thickness (or stacking) directions Z or 3 are the same in both frameworks. The (12) directions refer to the Cartesian framework defined by the averaged warp (1) and weft (2) tow directions. The (XY) directions refer to the testing machines system of axis (X=tensile direction).

2.2.1. Microstructure generation

The plain weave microstructure was ideally reproduced with the general characteristics of the real material (Fig. 3) as provided by the weaving company or as measured. The 2D plane view of the REV is a square of about $7.94mm$ in size and consists of four intertwined identical tows surrounded by the matrix. Optical microscopy was used to characterize tows by a cross section with an elliptic shape of $3.26mm \times 0.22mm$. The overall fiber volume fraction is $V_f = 0.47$. It should be noted that the plain weave balanced fabric creates a periodic distribution of full resin pockets of average square size equal to $0.7^2 = 0.49mm^2$. In the tows the fiber volume fraction is $V_f^{tows} = 0.7$.

The 3D modelling of the REV was performed using Solidworks® and is shown in Figure 3 which clearly shows the distribution of materials within sections at different depths of the fabric lamina following the tow undulations. The tows are shown here with the same 'color' but in the material description under CraFT, two sets of properties are given with same mechanical properties but in different orders, so as to take into account their different orientations. The real laminate composite is simulated as a superposition of nine plies. For this, two additional layers (one pixel size) on the bottom and up surfaces were considered as having null mechanical properties. Each ply was divided into 9 pixels along direction 3 which gives 83 pixels for the whole discretization along this direction. In the (12) plane, a unit cell was made up of $443 \times 443$ pixels which conserves the pattern's real aspect ratio and gives a resolution of about $7.94/443 \approx 18 \, \mu m/px$.

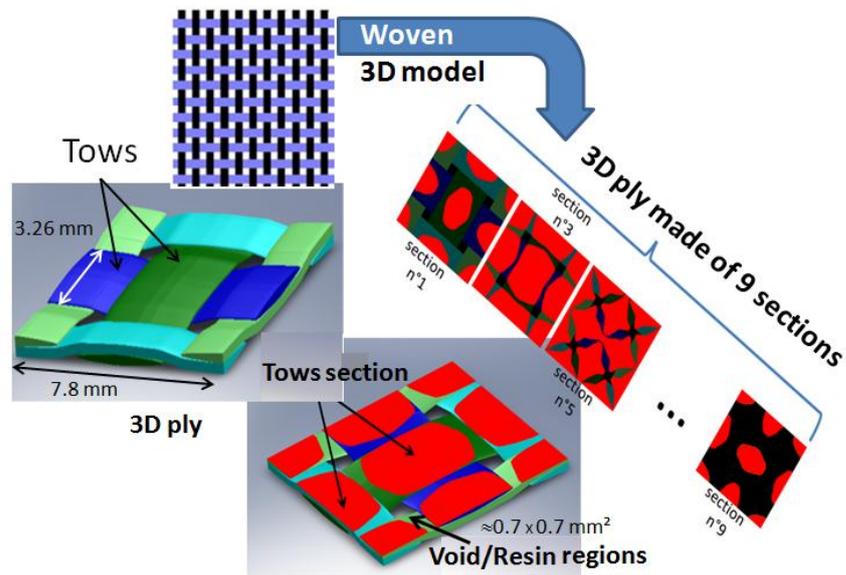

Figure 3: 3D numerical REV corresponding to the real woven and material distribution for different slices.

2.2.2. The mechanical properties at the meso-scale.

Once this geometrical description of the REV has been made, mechanical properties must be attributed to each "phase" of the material. Only the elastic behavior is considered for each phase and Table 1 lists the values of all relevant moduli. The resin was a newly referenced thermoplastic (Elium® - Arkema) considered to be homogeneous and isotropic and whose properties were characterized experimentally

(using Dynamic Mechanical Analysis-DMA for the elastic modulus determination) or from a well-established data base (Poisson coefficient for acrylic material). E-Glass fibers with well known mechanical properties and diameters in the $15-20\,\mu m$ range were used to perform the tows. Each tow was modelled as a unidirectional composite whose mechanical behavior can therefore be described with 5 independent elastic constants, for example $E_{11}$, $E_{22}$ $\nu_{12}$, $G_{12}$ and $G_{23}$ (1 is the average fiber direction in the tow). Many micromechanical models are available to calculate these homogenized elastic constants, for instance the Chamis model [Chamis, 1983], the Hopkins and Chamis model (HC) [Hopkins and Chamis, 1985] and the Composite Cylinder Assemblage model (CCA) [Rosen, 1970]. These models give more or less the same results but it should be noted that only the CCA model provides the $G_{23}$ value (see table 1). For the simulations shown in this paper we used the Chamis model for $E_{11}$, $E_{22}$ $\nu_{12}$ and $G_{12}$, and the CCA model for $G_{23}$. However, we checked that the conclusions obtained during our study remain qualitatively and quantitatively true even if other micromechanical models were used to evaluate the elastic constants in the tows.

Table 1: Mechanical properties of the different phases of the studied composite (1 = average tow direction).

|  |  | $E$ (GPa) |  | $\nu$ |  |  |
|---|---|---|---|---|---|---|
| Elium® resin |  | 3.6 |  | 0.37 |  |  |
| E-Glass fiber |  | 73 |  | 0.22 |  |  |
| Tow | Model | $E_{11}$ (GPa) | $E_{22}$ (GPa) | $\nu_{12}$ | $G_{12}$ (GPa) | $G_{23}$ (GPa) |
| $V_f^{tows}=0.7$ | Chamis | 52.2 | 17.6 | 0.265 | 6.57 | Not provided |
|  | HC | 52.2 | 15.3 | 0.265 | 5.71 | Not provided |
|  | CCA | 52.2 | 19,1 | 0.259 | 6.01 | 7.17 |

The variations of elastic constants expressed in the (1, 2, 3) coordinate system due to the fiber undulation effect were not taken into account. To check the validity of this approximation, we modelled the tow undulation using a sinusoidal function and found the variations of $E_{11}$ to be less than 5%. The elastic constants in the tows are indicated in table 1 for a tow along the 1 direction.

2.2.3. In-plane shear loading.

As mentioned in the introduction, in-plane shear loading conditions lead to the harshest conditions being applied to the fiber/matrix interface. In the (XY) plane, a tensile load was applied in the X-direction, at ±45° from the warp (1) and weft (2) fabric directions. The following plane stress tensor was applied:

$$\sigma = \begin{bmatrix} \sigma_\ell & 0 \\ 0 & 0 \end{bmatrix}_{(X,Y)} = \begin{bmatrix} \sigma_\ell/2 & \sigma_\ell/2 \\ \sigma_\ell/2 & \sigma_\ell/2 \end{bmatrix}_{(1,2)} \quad (5)$$

where $\sigma_\ell$ was considered to be the average loading stress imposed on the whole REV in CraFT software. Our goal in section 4 will be to analyse the stress field at the mesoscale which is linked to this uniform constant stress at the macroscale. It is evident from the tensor defined in (5) that in the direction of a tow (1 for example), the longitudinal stress $\sigma_{11}$ derives only from the fiber response whereas both the shear stress $\sigma_{12}$ and transverse stress $\sigma_{22}$ "work" in order to introduce fiber/matrix debonding.

In the following, we simulated the behavior of the above described composite in the case of tensile tests performed with warp and weft tows oriented at ±45° with respect to the tensile direction (ASTM D3518 or NF ISO 14129) and compared the simulated results to experimental ones. For both sections 3 and 4, all quantities with upperscripts M and m will denote values corresponding respectively to the Macroscale and the mesoscale.

3. **Analysis of strain fields**

Qualitatively, it is possible to compare the mesoscopic strain field $\varepsilon_{XX}^m$ obtained by CraFT simulation (Fig.4a) and by experimental DIC measurements (Fig.4b) using ARAMIS software. The X-axis denotes the tensile direction, while the 1 and 2 directions are correspond to the mutually orthogonal axis of the tows. It is clear that the exact same pattern was obtained for strain distribution which displays a strong correlation with the fabric geometry. The small squared areas of high strains was found to be located in the rich resin zones, those of small strains, in the rich fiber zone (interlaced tows corresponding to regions 1 and 2 in Fig.4a). The areas with intermediate strains (bands surrounding the interlaced regions and forming a cross pattern) correspond to the undulation nodes of the tows where the resin forms strong junctions between the laminates. On this qualitative comparison, we can also highlight the important difference in spatial strain resolution between the simulation (highly resolved spatial resolution set as 1/Np where Np=443 is the number of pixels chosen for the REV size) and the DIC measurements. The best resolution achieved for the latter was about one fifteenth the VER size ([Boufaida, 2015], see also Fig.5a). The ratio of simulation versus experimental resolution was found to be within a factor of 30.

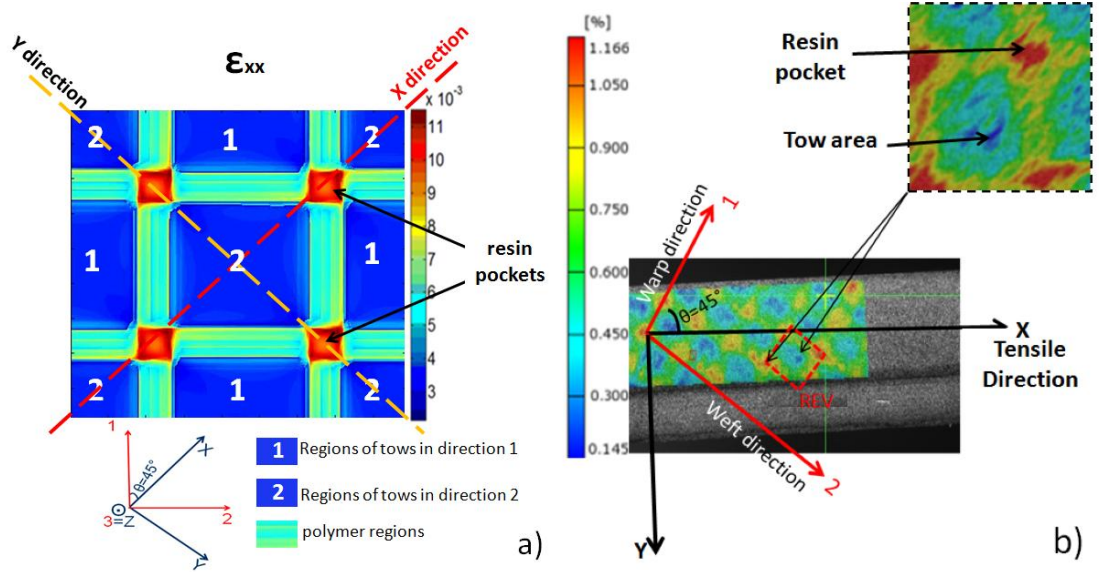

Figure 4: a) Numerical strain field $\varepsilon_{XX}$ (the 1 and 2 squares indicate the warp and weft tows regions); b) Experimental DIC strain field $\varepsilon_{XX}$.

Quantitatively, we calculated the Macroscopic elastic moduli $E_{XX}^M$ and $G_{12}^M$ from the averaged strain distribution over the simulation REV. The Macroscopic applied stress was $\sigma_{XX}^M = 50\ MPa$. We found $\langle \varepsilon_{XX} \rangle_{REV} = 0.00358$ and $\langle \varepsilon_{12} \rangle_{REV} = 0.0026$. The shear modulus must be calculated as $G_{12}^M = \sigma_{12}^M / 2\langle \varepsilon_{12} \rangle_{REV} = \sigma_{XX}^M / 4\langle \varepsilon_{12} \rangle_{REV}$ as a result of the 45° rotation angle between the (XY) and (12) axis systems (eq. (5)). We naturally obtained $E_{XX}^M = \sigma_{XX}^M / \langle \varepsilon_{XX} \rangle_{REV}$. Table 2 gives these results (1$^{st}$ row) along with those obtained using the above-mentioned micromechanical models (2$^{nd}$ row) applied to an equivalent laminate made up of three layers:

i) Two UD (UniDirectional) composite layers respectively oriented at $+45°$ and $-45°$ with fiber volume fraction corresponding to that of the tows ($V_f^{tows} = 0.7$) and

ii) a layer made of pure resin, which is added to account for the resin pockets. The relative thickness of these layers was chosen to comply with the overall composite fiber volume fraction ($V_f = 0.47$).

For the CraFT simulation, the in-plane elastic constants in the two composite layers were calculated using the Chamis model (see table 1). The results were found to be very close for $E_{XX}^M$ (within 3%) and for the $G_{12}^M$ values corresponding to the CraFT simulation and those calculated with the equivalent UD laminate approximation.

Table 2: Comparison of Macroscopic elastic moduli $E_{XX}^M$ and $G_{12}^M$

|  | $E_{XX}^M = \dfrac{\sigma_{XX}^M}{\langle \varepsilon_{XX} \rangle_{REV}}$ | $G_{12}^M = \dfrac{\sigma_{12}^M}{2\langle \varepsilon_{12} \rangle_{REV}}$ |
|---|---|---|
| CRAFT simulation | 13.94 GPa | 4.82 GPa |
| Equivalent UD laminate | 14.40 GPa | 4.82 GPa |
| Experimental values | 14.00 GPa (DMA) | 3.41 GPa (tensile test $\pm 45°$) |

Table 2 also gives experimental results (3$^{rd}$ row). The experimental value of $G_{12}^M$ obtained from the tensile curve was found to be 30% lower than the models. The discrepancy obviously came from an experimental under-estimation of our measurement. The stress-strain curve obtained experimentally during $\pm 45°$ tensile testing with strains measured using 3D-DIC was not linear [Boufaida et al., 2015], even in the strain range where the $G_{12}^M$ shear modulus needed to be evaluated according to the ASTM D3518 standard i.e. for $2\varepsilon_{12}^M \in [0.001 \quad 0.005]$. This suggests that the resin had a noticeable visco-elastic effect in this kind of stress state. As a result, the elastic behavior needs to be looked for at very early times of the tensile test which is unfortunately when the machine response can be smoothed by the feedback loop resulting in a lower estimation of the Young Modulus. This fact was identified when measuring the instantaneous modulus of semi-crystalline polymers - a visco-elastic physical model needs to be applied to the whole tensile curve to obtain a good estimation of the modulus [Blaise et al., 2016]. Therefore, to avoid this drawback, the experimental measurement of $E_{XX}^M$ provided in Table 2 was obtained through DMA performed on a Bose 3300N fatigue machine. DMA enables the obtention of the "true" elastic modulus of a visco-elastic material (real part of the complex modulus). The agreement between the $E_{XX}^M$ values obtained by the CraFT simulation and with the DMA experiment was found to be very good (within 0.5%). Proof of the discrepancy in the $G_{12}^M$ values attributed to visco-elasticity can be seen in the following calculations - for the $\pm 45°$ fiber orientation, the relation between $E_{XX}^M$ and the composite elastic constants related to the $(1,2,3)$ coordinate system was found to be: $\frac{1}{E_{XX}^M} = \frac{1}{4}\left[\frac{1}{G_{12}^M} + \frac{2(1-\nu_{12}^M)}{E_{11}^M}\right]$. $E_{11}^M$ is the Macroscopic Young modulus along the warp or equivalent weft direction and $\nu_{12}^M$ is the Macroscopic plain weave composite Poisson's ratio. These quantities must not be confused with $E_{11}$ and $\nu_{12}$ from table 1 which are local values in the tows.

$G_{12}^M$ can then be extracted from the previous equation:

$$G_{12}^M = \frac{E_{11}^M E_{XX}^M}{4E_{11}^M - 2(1-\nu_{12}^M)E_{XX}^M} \tag{6}$$

As $E_{XX}^M$, $E_{11}^M$ was measured by DMA: $E_{11}^M = 23.7\, MPa$. $\nu_{12}^M$ can be evaluated with the equivalent UD laminate model presented above: $\nu_{12}^M = 0.15$. The dependence of $G_{12}^M$ on $\nu_{12}^M$ in eq.6 is very small which can be verified. The $G_{12}^M$ value that can be calculated with eq.8 was therefore obtained with realistic elastic constant measurements, in other words without any bias due to visco-elasticity. Eq.6 gives the new experimental value of $G_{12}^M = 4.67\, GPa$. The difference with the value obtained with the CraFT simulation ($G_{12}^M = 4.82\, GPa$) is now only about 3%. This confirms that the discrepancy between the $G_{12}^M$ value measured during $\pm 45°$ tensile testing and the value obtained with the CraFT solver can indeed be attributed to visco-elastic effects which are not accounted for in the numerical approach.

Figure 5b shows the three $\varepsilon_{XX}^{meso}$ strain profiles obtained theoretically (solid lines) and experimentally (dashed line). There was clear evidence of the difference in resolution in the 2D patterns (Fig.5a) and also in the curves of Fig.5b. The DIC measurements therefore correspond to a kind of averaging of the strain distribution over non negligible zones. It was therefore impossible to get a good resolution of strong strain gradients and to capture strong strain variations which were still present because of the transitions from rigid tows to the softer resin material. As a result, the highly deformed zones associated with the resin pockets appeared to be magnified to some extent but a strong quantitative validation of both results can be obtained by looking at the profiles of Fig.5b. The experimental and theoretical strain profiles were found to be very similar. Of course the experimental curve appears very noisy with an enlarged base-peak as a result of the low measurement resolution. The theoretical profiles were once again nearly symmetrical but there were two peaks at the tow/resin interface which could not be present on the experimental curve for the same reason. This suggests that the maximum strains are located exactly at the interface in the resin phase. The ratio between the strains in the fiber rich and resin rich regions (Fig. 5a) was found to be approximately 3 both experimentally and by simulation. In an attempt to reproduce the experimental results more closely, we considered the highly resolved strain patterns to be real objective measurements and tried to reproduce the spatial filtering operated when using our 3D-DIC system. The whole procedure is detailed in the appendix. In shortly, the highly resolved strain field was integrated to obtain the displacement field (in the small displacement-gradient theory framework). For this, the heterogeneous component of the strain (considered as periodic over the REV) was transformed in the Fourier domain to produce the Fourier transform of the displacement which was back-transformed in the real domain. From this highly-resolved displacement field, we used an averaging procedure to simulate the behavior of the Aramis software and obtain the transformation gradient tensor at a smaller resolution.

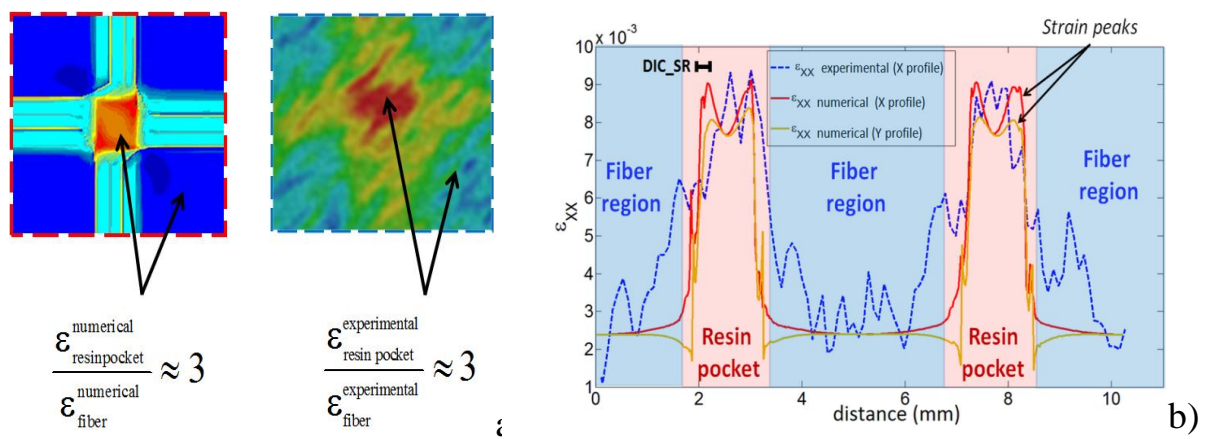

Figure 5 : a) Numerical and experimental strain patterns of $\varepsilon_{XX}^{meso}$ around a resin pocket; b) Numerical and experimental profiles of the $\varepsilon_{XX}^{meso}$ strain across a REV along profiles Nr 1,2.

Figure 6 (or A-2 in the Appendix) shows a new strain pattern obtained numerically after lowering the resolution and the corresponding numerical and experimental profiles along directions 1 or 2. The peaks found at the resin-tow interfaces can be seen to have disappeared while the strain profiles now match the experimental ones much better (apparent enlargement of the peak bottom).

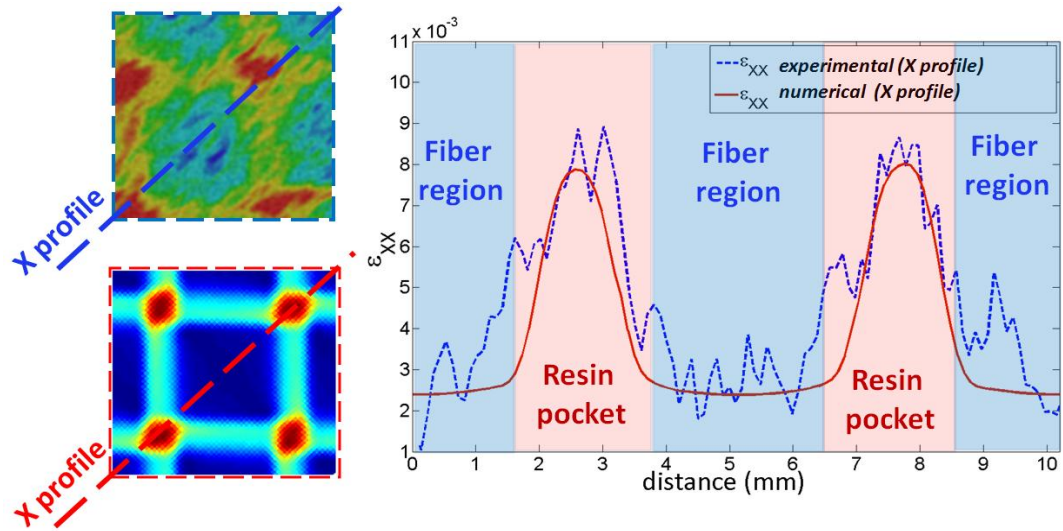

Figure 6 : Numerical and experimental profiles of the $\varepsilon_{XX}^{meso}$ strain along the 45°lines after applying a smoothing procedure to CRAFT results.

## 4. Stress fields analysis

Following on from the validation provided by numerical and experimental comparisons of strain fields, the stress field obtained numerically was then investigated. Within the scope of damage initiation through fiber-matrix debonding, experimental investigations based on X-ray tomography imaging were found to confirm the results.

For a woven fabric composite material, the stress field is totally governed by the mesoscale resulting from the textile geometry. The average stress applied to a micrometric REV (inside the tows) does not correspond to the average macroscopic stress but to the mesoscopic stress in the region where the micrometric REV is taken. The detailed stress field structure at the mesoscale is required to accurately analyze the damage phenomenon related to the fiber-matrix debonding in a fabric. In the following, the stress field is analyzed for the tows with their fibers oriented in the 2-direction (central square of Fig.4a or Fig. 7 below). We focused on the $\sigma_{12}^m$ and $\sigma_{11}^m$ stress components which were the most likely be able to test the fiber-matrix resistance because the $\sigma_{22}^m$ component of the meso-stress tensor would be mainly supported by the fibers (as in a UD composite with applied load along the fibers). Figure 7 shows the $\sigma_{12}^m$ stress map obtained from CRAFT simulations which is given here in the fiber framework $R_1(O,1,2,3)$ and for a macroscopic applied stress $\sigma_{XX}^M = 50\ MPa$. Highly-sheared regions appeared to be located in the tows at the tow-resin interface. The contrast in shear stress was of about factor 5 when crossing this interface. This plane shear test effectively produces a high shear stress level at the mesoscale in order to solicit the matrix-tow and fiber-matrix adhesion. The plot of Figure 8 corresponds to the profile of the $\sigma_{12}^m$ stress along the tensile direction. The tow can be seen to be highly stressed on its edges (stress drop again of a factor 5-6) and the resin in the pocket zones was subjected to strong stress gradients. Indeed the shear stress was about 5 MPa at the interface and 15 MPa in the center as compared with a macroscopic stress of $\sigma_{12}^M = \sigma_{XX}^M/2 = 25\ MPa$ which was obviously mainly supported by the tows (Figures 7 and 8). In the fiber region (at the centre of the REV of Fig.8a), and in view of the fiber/matrix bond resistance estimation, the ratio of the max-min shear stress $\sigma_{12}^{max}/\sigma_{12}^{min}$ was approximately 1.3 which is not particularly strong.

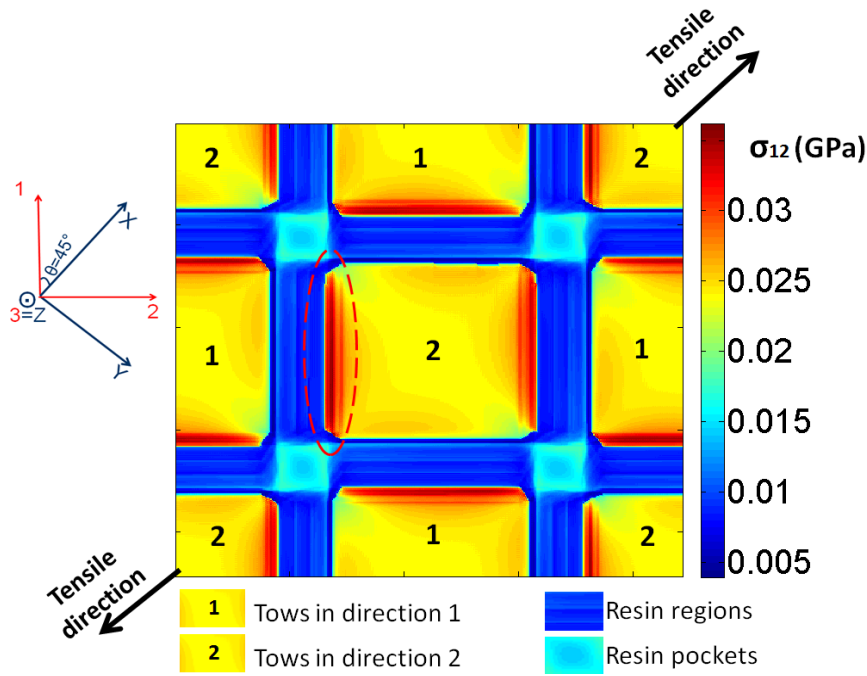

Figure 7: $\sigma_{12}^m$ stress map (CRAFT Simulation for $\sigma_{XX}^M = 50\ MPa$)

The $\sigma_{11}^m$ stress map (Figure 9-a) provides additional information which shows that the regions of highest stress were very small and located in the corners of the tows. There, the mechanical solicitation was oriented perpendicularly to the direction of the fibers in the tow and could (i) introduce fiber-matrix debonding with further cracks propagation and (ii) create important damage in the form of cavities in the pocket resin.

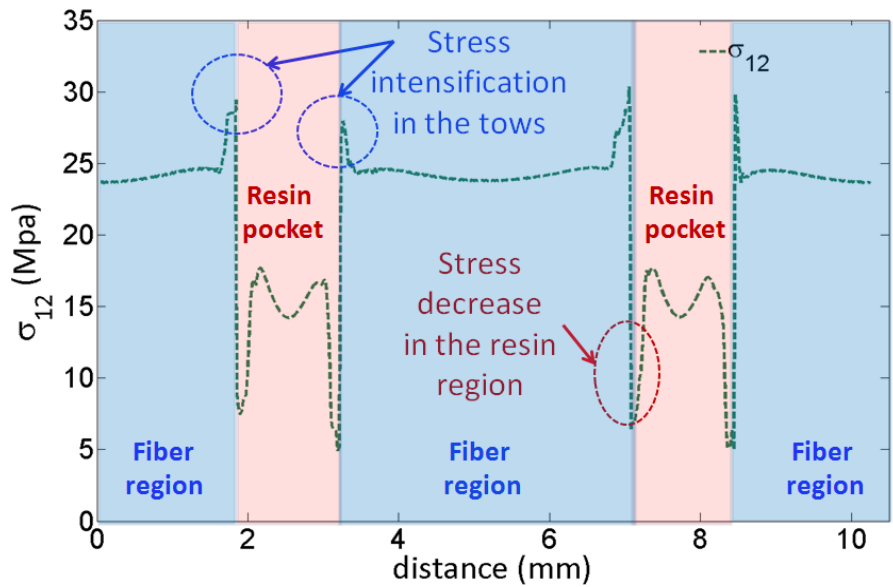

Figure 8: $\sigma_{12}^m$ stress profile along the tensile direction of Figure 9.

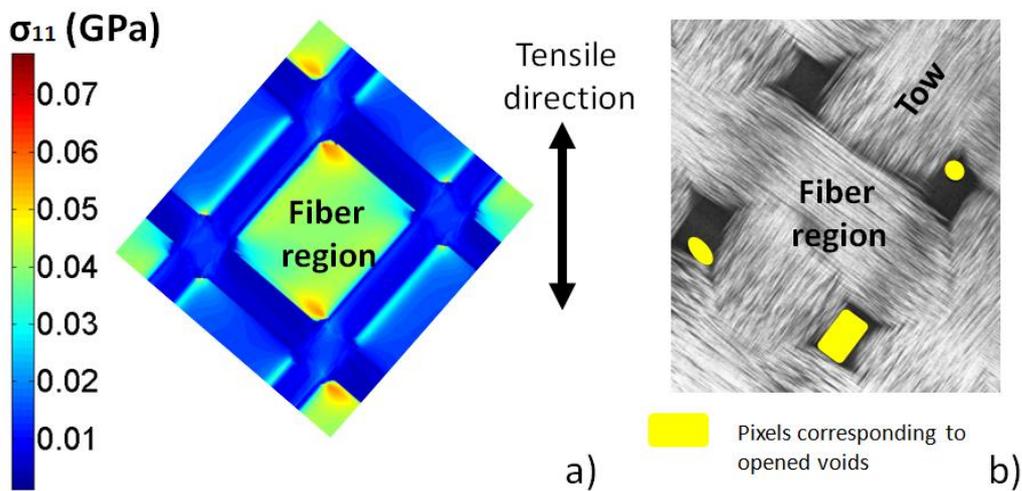

Figure 9: a) $\sigma_{11}^m$ stress map (CRAFT Simulation for $\sigma_{XX}^M = 50\ MPa$)
b) X-ray tomography cliché on real specimen.

This is precisely what can be observed in Figure 9-b which presents an X-ray tomography snapshot of the real specimen tested in such plane shear test. Moreover, heat build-up experiments reported in Boufaida et al. (2015 – Figs 5a and 5b) and consisting of the detection of a change in thermomechanical regime under fatigue excitations showed that the typical stress initiating the so-obtained damage yield is of the order of $25\ MPa$. In fact, this value drives the $\sigma_{XX}^{M}$ value of 50MPa taken for the simulations presented here which give a Von Mises stress in the resin pocket in the 28 Mpa – 33 MPa range (depending on whether the center or the edges of the zone are considered). It is of the order of 68 MPa in the corner of the crossing tows. This value is much greater than the stress at rupture of acrylic resin which is of the order of $\sigma^r \approx 44 MPa$. All these numerical and experimental results therefore match very reasonably and show that the CRAFT tool and associate spectral solver are very efficient and have interesting prospects for engineering applications on composite materials.

Lastly it should be noted that in the fiber region, the ratio of max/min values $\sigma_{11}^{m,\max}/\sigma_{11}^{m,\min}$ was of about 2. This is quite a large value and can be seen as a strong stress concentration at the mesostructure scale although it occurred in the same microstructural object, namely the tow. Following on from this observation, future study of the behavior of the heterogeneous tow with respect to damaging phenomena could be an interesting research direction.

## 5. Conclusion

This study presented numerical simulation results of the mechanical behavior of a woven fabric composite in the pure elastic case. Despite the complex geometry and strong heterogeneous properties of the Representative Elementary Volume (acrylic resin, Glass fibers arranged in warp and weft tows at 90°), the results serve as an explanation for in-depth experimental observation in terms of strains (direct comparison with full-field DIC measurements) and stresses (indirect comparison through the characterization of damage initiation). All data for the simulations was provided by independent characterizations performed by the authors and introduced, in some cases, into micromechanical models to determine properties at the mesoscale. It would be a good idea for future research based on this spectral solver to deal with modelling of the tows themselves in the stress field configuration obtained at the mesoscale to improve the overall numerical predictions.

**Acknowledgments:** The authors wish to thank the Région Lorraine for financial support and PhD grant, Alain Gérard, technician at LEMTA who provided technical assistance for the experiments, Hervé Moulinec (LMA) for a fruitful introduction to CraFT and guidance in using it.